\documentclass[11pt,a4paper]{article}
\usepackage[utf8]{inputenc}
\usepackage[english]{babel}
\usepackage[sorting=none]{biblatex}
\DefineBibliographyStrings{english}{references = {References}}

\usepackage{authblk}

\usepackage{color}
\usepackage[colorlinks=true, linkcolor=red, citecolor=blue]{hyperref}
\usepackage{amsmath}
\usepackage{amssymb}
\usepackage{amsthm}
\usepackage{sectsty}
\chapterfont{\centering\fontsize{17}{0pt}\selectfont}
\sectionfont{\fontsize{14}{0pt}\selectfont}
\subsectionfont{\fontsize{13}{0pt}\selectfont}
\subsubsectionfont{\fontsize{12}{0pt}\selectfont}
\usepackage{etoolbox}

\usepackage{array}
\usepackage{mathtools}
\usepackage{empheq}
\usepackage[toc,page]{appendix}
\usepackage{tabu}
\usepackage{tabularx}
\usepackage{booktabs}

\usepackage{graphicx}

\usepackage{subfig}
\usepackage{enumerate}

\usepackage{commath}

\usepackage{mathptmx}
\usepackage{setspace} 
\usepackage{stack} 
\usepackage{eqnarray}
\usepackage{tocbibind}
\usepackage[top=1in,bottom=1in,left=1.5in,right=1in]{geometry}
\usepackage{float}
\usepackage{caption}

\usepackage{tikz}
\usetikzlibrary{arrows,shapes,trees,intersections,graphs,positioning,fit,automata,calc,chains,decorations,shadows,backgrounds}
\tikzstyle{rec} = [rectangle, rounded corners, minimum width=3cm, minimum height=1cm, text centered,draw=black, fill=red!10]
\tikzstyle{arrow} = [thick,->,>=stealth]
\tikzstyle{arrowbar} = [thick,->,>=stealth]
\tikzstyle{recdot} = [draw=pink, dashed, rectangle,rounded corners,minimum width=3cm,minimum height=1cm,text centered,fill=blue!2]
\tikzstyle{smallbox} = [rectangle, rounded corners, draw=black, text centered]

\tikzstyle{mycirc} = [circle,draw=black,ultra thick]

\usepackage{pgfplots} 
\pgfplotsset{compat=1.12} 

\usepackage{csquotes}
\usepackage{comment}
\usepackage{enumerate}
\usepackage{colortbl}

\usepackage[T1]{fontenc}

\author[1,2]{Christian Nwachioma\thanks{chris.nwachioma@gmail.com, }}
\author[1]{Farida Tahir\thanks{farida\_tahir@comsats.edu.pk}}
\affil{COMSATS Institute of Information Technology, Islamabad.}
\affil[2]{National Mathematical Centre, Abuja}
\date{\today}
\newcommand{\apvec}[2]{\underset{\fontsize{3}{2}\selectfont {#1}}{#2}{}}
\title{Path Equations in Absolute Parallelism Space}
\begin{document}
    \maketitle
     \section*{Abstract}
     Riemannian and Absolute Parallelism (AP) geometries are discussed. A lavish treatment of path equations in the AP-space using the Bazanski-type Lagrangian is presented; We write down an expression that is absolutely conserved along a curve in the AP-space and show that it gives the same path equations as obtained using the conventional Lagrangian formalism. We attribute physical interpretations to the resultant path equations and we show that the spin-torsion interaction is an inherent property of the AP-space.

     \section{Introduction\label{one}}
     The Remannian metric tensor is the building block of all Riemannian geometric objects. It is spacetime dependent and represents the gravitational potential, which has been employed to describe gravity as a \textit{curvature field} in spacetime. The field approach is germane whenever we have a mass moving at relativistic speed. This is necessary because for instance, as two masses move relative to each other at high-enough speed, the radial distance between them will contract\cite{Arnab1998} making the Newton inverse-squared approach inappropriate as one mass cannot communicate its relative separation to the other mass at infinite speed since speed in itself is bounded above by light's speed. Also, the Absolute Parallelism (AP) space is discussed in this text: the basic vector of this space is a 4-dimensional parallelization vector or \textit{tetrad}; details about these spaces will be presented in bits especially whenever the mathematical tools to make them lucid arise.
     \subsection{Absolute and Covariant Derivative}
     One may wish to determine how a vector field varies along a curve by differentiating the vector with respect to the parameter $s$, characterizing the curve or with respect to coordinate axes $x^\alpha(s)$. The former is called absolute derivative of the vector field and the latter is called the covariant derivative. Given an arbitrary vector field $A^\mu(x^\nu)$ that is dependent on the coordinates, its absolute derivative is denoted by $DA^\mu/ds$ whereas its covariant derivative is denoted by $DA^\mu/dx^\nu$. The absolute and the covariant derivative of tensor field with components $A^\alpha$, are related as follows\cite{Dalarsson2005}.
     \begin{equation}
     \begin{split}
     {DA^\alpha\over ds}={DA^\alpha\over dx^\nu}{dx^\nu\over ds}\label{abs1}
     \end{split}
     \end{equation}
     In Euclidean space, particularly the Cartesian coordinate system, the attributes of a vector upon transportation from a point $\mathbf{x}$ to nearby point $\mathbf{x}+\mathbf{dx}$ is independent of the path taken. Thus, in this space, the covariant differential $DA^\mu$ of a vector $A^\mu$ is equivalent to our classical intuition of differential, that is:
     \begin{equation}
     \begin{split}
     DA^\alpha=dA^\alpha\label{abs2}
     \end{split}
     \end{equation} 
     However, on a curved manifold, the attributes of a vector upon parallel transportation is path dependent. In order to make a parallel transported vector path independent on this manifold, a certain property of the manifold has to keep track of the path traversed by a vector in order to effectively ensure the vector is same at a particular point on the manifold irrespective of the path taken to arrive at the point. This property of the manifold is what is called \textit{connection} and we denote it by $\Gamma$. Based on this, our classical intuition of differential of a vector transforms to become\cite{Lovelock1989}:
     \begin{equation}
     \begin{split}
     dA^\alpha\rightarrow dA^\alpha+\Gamma^\alpha{}_{\mu\nu}A^\mu dx^\nu=DA^\alpha\label{abs3}
     \end{split}
     \end{equation}
     Replacing the $DA^\alpha$ in Eq.\eqref{abs1} by the result of Eq.\eqref{abs3}, we have:
     \begin{equation}
     \begin{split}
     {DA^\alpha\over ds}&={dA^\alpha+\Gamma^\alpha{}_{\mu\nu}A^\mu dx^\nu\over dx^\nu}\enspace{dx^\nu\over ds}\\
     &=\big({dA^\alpha\over dx^\nu}+\Gamma^\alpha{}_{\mu\nu}A^\mu\big){dx^\nu\over ds}\\
     &={dA^\nu\over ds}+\Gamma^\alpha{}_{\mu\nu}A^\mu\dot{x}^\nu\label{abs4}
     \end{split}
     \end{equation}
     Eq.\eqref{abs4} is the absolute derivative of the vector field component, $A^\alpha(x^\nu)$; comparing Eq.\eqref{abs1} and Eq.\eqref{abs4}, it easy to see that the term in parenthesis in the second line of Eq.\eqref{abs4}, is the covariant derivative stated as follows.
     \begin{equation}
     \begin{split}
     D_\nu A^\alpha=A^\alpha{}_{,\nu}+\Gamma^\alpha{}_{\mu\nu}A^\mu\label{abs5}
     \end{split}
     \end{equation}
     Where $A^\alpha{}_{,\nu}$ implies ordinary differentiation of $A^\alpha$ with respect to $x^\nu$ : the type done in Euclidean space. For a $(0, 1)$ tensor field, the indices will be oriented accordingly and the sign affix to the \textit{connection} will be negative. 
     
\begin{figure}[H]
         \centering
         \includegraphics[scale=0.5]{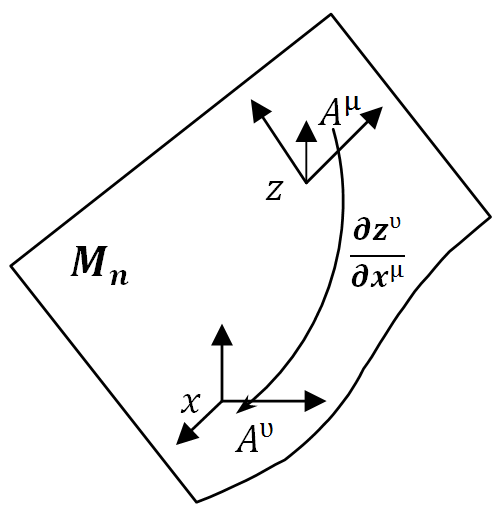}
         \caption{Coordinate Transformation on a Manifold, $M_n$\label{coordtrans}}
        \end{figure}

     As shown in \figref{coordtrans}, a vector field may be transformed from one system of coordinate to another. The transformation rule of a (1, 0) tensor transforming from coordinate $z^\mu$ to coordinate $x^\mu$ is given as follows\cite{Lovett2010}.
     \begin{equation}
     \begin{split}
     A^\mu(z)\rightarrow{\partial x^\nu\over\partial z^\mu}A^\mu(z)=A^\nu(x)\label{abs6}
     \end{split}
     \end{equation}
     The subtle difference in the transformation rule for a (0, 1) tensor should be noted in the following.
     \begin{equation}
     \begin{split}
     A_\mu(z)\rightarrow{\partial z^\mu\over\partial x^\nu}A_\mu(z)=A_\nu(x)\label{abs7}
     \end{split}
     \end{equation}
     It should be noted that the transformations at Eq.\eqref{abs6} and Eq.\eqref{abs7} are such that physical quantities such as scalars and indeed the tensor fields themselves remain chart independent or invariant as shown in Eq.\eqref{abs8}.
     \begin{equation}
     \begin{split}
     A^2=A^\nu A_\nu\rightarrow{\partial x^\nu\over\partial z^\mu}A^\mu{\partial z^\mu\over\partial x^\nu}A_\mu=A^\mu A_\mu = A^2\label{abs8}
     \end{split}
     \end{equation}
        
        Th \textit{connection}, since it functions to restore covariance to a parallel-transported vector, is itself, not a tensor. This is evident as given in Eq.\eqref{abs9}, in the way it transforms upon coordinate change in a curved space\footnote{ A manifold with a metric space structure, $(\mathbb{M},\mathbf{g})$, is said to be flat if a certain coordinate $\{f(x)\}$ can be found on the manifold where if the connection $\Gamma(x)$ is transformed to, it gives a vanishing of all the connection components in the coordinate $\{f(x)\}$; on this kind of manifold, the curvature tensor field $\mathbf{R}(f(x))$ will clearly yield zero result. Therefore, the curvature field being a tensor will always give zero in any coordinate system defined on the manifold including those for which the connection coefficient aren't all vanishing. We present the following as a mathematical definition of a flat space, where the arrow implies a coordinate transformation, $\mathit{f}$.
            \begin{equation}
            0\neq\Gamma^\alpha_{\mu\nu}(\mathbf{x})\longrightarrow\Gamma^{\alpha^\prime}_{\mu^\prime\nu^\prime}(\mathbf{\mathit{f}(x)})=0\implies \mathbf{R(\mathit{f}(x))}=0\therefore \mathbf{R(x)}=0\enspace\forall\enspace \{\mathbf{z}\}\in(\mathbb{M},\mathbf{g})  
            \end{equation}
            Where $\{\mathbf{z}\}$ represents different coordinate systems on the manifold.
            
            A curved spacetime on the hand, is a metric space defined on a manifold upon which it is impossible to find a coordinate system where the components of the connection function all vanish identically and the connection is heavily chart-dependent at every space point of the manifold. In other words, we cannot find a coordinate system in a curved spacetime where the connection coefficients are zero or constant. This has the consequence that the curvature tensor, which is written completely in terms of the connection is never zero in curved spacetime\cite{Nwachioma2016}.}.
        \begin{equation}
        \begin{split}
        \Gamma^\epsilon{}_{\mu\nu}(\mathbf{x})=\Gamma^\alpha{}_{\rho\lambda}(\mathbf{z}){\partial z^\rho\over\partial x^\mu}{\partial z^\lambda\over\partial x^\nu}{\partial x^\epsilon\over\partial z^\alpha}+{\partial^2z^\alpha\over\partial x^\mu\partial x^\nu}{\partial x^\epsilon\over\partial z^\alpha}\label{abs9}
        \end{split}    
        \end{equation}
        The second term containing the nonlinear partial derivative in Eq.\eqref{abs9} is the term, which has warranted us to declassified the \textit{connection} as a tensor since a tensor should transform as a linear or multilinear map depending on the tensor order.
        
        Furthermore, absolute and or covariant differentiation can be extended for a tensor of arbitrary rank. We have seen at Eq.\eqref{abs5}, that absolute (or covariant) derivative of a rank $1$ tensor contains a single \textit{connection}, analytically, we posit that absolute (or covariant) derivative of a tensor of rank $n$ will contain $n$ number of \textit{connections}. More so, while covariant derivative of a tensor field of rank $n$ yields a tensor of rank $n+1$, absolute derivative of a tensor of rank $n$, leaves the rank unchanged since absolute derivative is with respect to a scalar parameter. In what follows, we show that the absolute (or covariant) derivative of a scalar or rank-$0$ tensor contains no \textit{connection}, which means that absolute (or covariant) differentiation and the corresponding ordinary differentiation are equivalent operations for a scalar function say, $\phi(x)$.
        \begin{equation}
        \begin{split}
        {D\over ds}(A^\mu A_\mu)&={D\phi\over ds}\\
        &={D\phi\over dx^\nu}{dx^\nu\over ds}\\
        &={D\over dx^\nu}(A^\mu A_\mu){dx^\nu\over ds}\\
        &=\big({DA^\mu\over dx^\nu}A_\mu+A^\mu{DA_\mu\over dx^\nu}\big){dx^\nu\over ds}\\
        &=\big((A^\mu{}_{,\nu}+\Gamma^\mu_{\alpha\nu}A^\alpha)A_\mu+A^\mu(A_{\mu,\nu}-\Gamma^\alpha_{\mu\nu}A_\alpha)\big){dx^\nu\over ds}\\
        &=(A^\mu{}_{,\nu}A_\mu+A^\mu A_{\mu,\nu}){dx^\nu\over ds}\\
        &={d\over dx^\nu}(A^\mu A_\mu){dx^\nu\over ds}\\
        &={d\phi\over dx^\nu}{dx^\nu\over ds}\\
        &={d\phi\over ds}\label{abs10}
        \end{split}
        \end{equation}
        Eq.\eqref{abs10} implies $D\phi/ds=d\phi/ds$ and $D\phi/dx^\nu=d\phi/dx^\nu$. As expected, since $\phi$ is a rank-$0$ tensor, it contains zero number of \textit{connection}. Also, it can be seen that the absolute derivative leaves the rank at zero while covariant derivative raises the rank from zero to one; in fact, the covariant derivative is easily seen to be a generalization of the gradient of a function. Physically, the above result implies a scalar will not be affected by spacetime curvsature.

        \subsection{Riemmanian Space}
        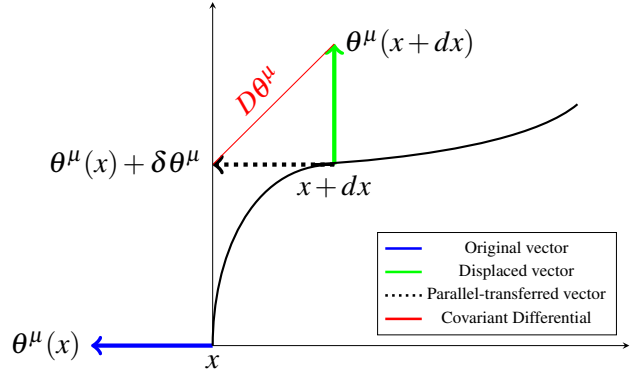
\begin{figure}[h!]
            \centering
            \begin{tikzpicture}[scale=0.8]
            \coordinate [label=below:${x}$] (A) at(0,0);
            \coordinate [label=left:$\theta^\mu(x)$](B) at (-2,0);
            
            \coordinate [label=below:${x+dx}$](C) at (2,3);
            \coordinate [label=right:$\theta^\mu(x+dx)$](D) at (2,5);
            
            \coordinate [label=left:${}$](E) at (2,3);
            \coordinate [label=left:$\theta^\mu(x)+\delta\theta^\mu$](F) at (0,3);
            
            \draw[blue, ultra thick,->] (A) -- (B);
            \draw[green, ultra thick,->] (C) -- (D);
            \draw[dotted,ultra thick,->] (E) -- (F);
            \draw[red,thin,-] (2,5) -- (0,3) node[pos=0.5,sloped,above]{$D\theta^\mu$};
            
            \draw[thick] (A) to [out=90,in=180] (C) to [out=180,in=220] (6,4); 
            
            \begin{axis}[
            legend pos=south east,
            xtick=\empty,
            ytick=\empty,
            axis y line=left,
            axis x line=bottom,
            legend style={font=\fontsize{1}{2}\selectfont}
            ]
            \addplot[blue,very thick] coordinates {(7,0)};
            \addplot[green,very thick] coordinates {(7,1)};
            \addplot[dotted,very thick] coordinates {(7,2)};
            \addplot[red,very thick] coordinates {(7,3)};
            \legend{\fontsize{8}{10}\selectfont Original vector,
                \fontsize{8}{10}\selectfont Displaced vector,
                \fontsize{8}{10}\selectfont Parallel-transferred vector,
                \fontsize{8}{10}\selectfont Covariant Differential};
            \end{axis}
            \end{tikzpicture} 
            \caption{A vector $\theta^\mu$ transported along a curve \label{parallelTransfer}} 
        \end{figure}   
        
        \figref{parallelTransfer} depicts the concept of vector transportation in a curved spacetime; a vector has been parallel transported from point $\mathbf{x}$ to point $\mathbf{x}+\mathbf{dx}$ and the same vector when displaced in parallel sense through a different route and then brought to the same point $\mathbf{x}+\mathbf{dx}$, will be found to differ from the first. But on this same manifold, invariance with respect to parallel transportation has been required of the fundamental building block of every geometric object on the manifold; this fundamental building block is the metric tensor, $g_{\mu\nu}$ \cite{Moshe2006,Wald1984}; this is equivalent mathematically, to the requirement that the metric be invariant with respect to a covariant differentiation, $D_\alpha$, that is:
        \begin{equation}
        \begin{split}
        0&=D_\alpha g_{\mu\nu}=g_{\mu\nu;\alpha}\label{eqn1}
        \end{split}
        \end{equation}
        Where the semicolon denote covariant differentiation with respect to the Riemannian connection, $\{^\epsilon_{\mu\nu}\}$. Employing the rule of covariant differentiation, Eq.\eqref{eqn1} can be written out more explicitly as follows.
        \begin{equation}
        \begin{split}
        0=g_{\mu\nu,\alpha}-\{^\epsilon_{\mu\alpha}\}g_{\epsilon\nu}-\{^\epsilon_{\nu\alpha}\}g_{\mu\epsilon}\label{eqn2}
        \end{split}
        \end{equation}
        Eq.\eqref{eqn2} clearly indicates that ordinary differentiation of the spacetime dependent metric tensor can be written as:
        \begin{equation}
        \begin{split}
        g_{\mu\nu,\alpha}=\{^\epsilon_{\mu\alpha}\}g_{\epsilon\nu}+\{^\epsilon_{\nu\alpha}\}g_{\mu\epsilon}\label{eqn3}
        \end{split}
        \end{equation}
        To make $\{^\epsilon_{\mu\alpha}\}$ a subject and write it in terms of the metric, the indices $\mu$, $\nu$, and $\alpha$ are alternated to get:
        \begin{equation}
        \begin{split}
        g_{\mu\nu,\alpha}=\{^\epsilon_{\mu\alpha}\}g_{\epsilon\nu}+\{^\epsilon_{\nu\alpha}\}g_{\mu\epsilon}\\
        g_{\mu\alpha,\nu}=\{^\epsilon_{\alpha\nu}\}g_{\mu\epsilon}+\{^\epsilon_{\mu\nu}\}g_{\nu\epsilon}\label{eqn3.1}\\
        g_{\nu\alpha,\mu}=\{^\epsilon_{\nu\mu}\}g_{\epsilon\alpha}+\{^\epsilon_{\alpha\mu}\}g_{\nu\epsilon}
        \end{split}
        \end{equation}
        It will be implausible to add up all terms of Eq.\eqref{eqn3.1} in order to make a connection the subject, instead and thanks to the symmetries in the lower indices of $\{^\epsilon_{\mu\alpha}\}$ and of $g_{\mu\nu}$, we can eliminate pairs that are equivalent. Thus, sticking a minus in any one line of Eq.\eqref{eqn3.1}, we have:
        \begin{equation}
        \begin{split}
        g_{\mu\alpha,\nu}+g_{\nu\alpha,\mu}-g_{\mu\nu,\alpha}=2g_{\epsilon\alpha}\{^\epsilon_{\mu\nu}\}\label{eqn3.2}
        \end{split}
        \end{equation}
        By the way, we have stuck the minus on the first line of Eq.\eqref{eqn3.1}. In Eq.\eqref{eqn3.2}, $g_{\epsilon\alpha}$ cannot act on $\{^\epsilon_{\mu\nu}\}$ because the slots are completely filled and because nonlinear term appears in the connection upon a coordinate transformation; so the other option to free $\{^\epsilon_{\mu\nu}\}$ and make it indeed a subject, is to detach from it, the $g_{\epsilon\alpha}$; and we have to be choosy on the choice of indices of the inverse metric because for instance $g^{\epsilon\rho}$ will not achieve the desired result, whereas $g^{\alpha\rho}$ will; because the former will give $\delta_\alpha^\rho$ still attached to $\{^\epsilon_{\mu\nu}\}$ and yet incapable of acting on it but the later will be good to go. Thus, multiplying through by $g^{\alpha\rho}$, we have:
        \begin{equation}
        \begin{split}
        g^{\alpha\rho}(g_{\mu\alpha,\nu}+g_{\nu\alpha,\mu}-g_{\mu\nu,\alpha})=2\delta^\rho_\epsilon\{^\epsilon_{\mu\nu}\}\label{eqn3.3}
        \end{split}
        \end{equation}
        Finally, $\{^\epsilon_{\mu\nu}\}$ is free and it's a subject written solely in terms of the metric as follows.
        \begin{equation}
        \begin{split}
        \{^\rho_{\mu\nu}\}=\frac{1}{2}g^{\alpha\rho}(g_{\mu\alpha,\nu}+g_{\nu\alpha,\mu}-g_{\mu\nu,\alpha})\label{eqn4}
        \end{split}
        \end{equation}
        
        The Riemannian connection at Eq.\eqref{eqn4} is often called the Christoffel symbol of the second kind; it is a non-tensor object, which accounts for invariance of vectors and higher other tensors in curved spacetime; it does this by adjusting appropriately to uphold the invariance expectation of a vector transported through a curved spacetime manifold as the following shows.
        \begin{equation}
        \begin{split}
        A_{\mu,\nu}-A_\alpha\{^\alpha_{\mu\nu}\}=D_\nu A_\mu\label{eqn5}
        \end{split}
        \end{equation}
        The above is the covariant derivative of a $(0, 1)$-tensor field and it is denoted by $D_\nu A_\mu$. A $(1, 0)$-tensor field will have its covariant derivative as $A^\alpha{}_{,\nu}+A^\mu\{^\alpha_{\mu\nu}\}=D_\nu A^\alpha$. As stated earlier, the covariant derivative of a rank $n$ tensor will result in a rank $n+1$ tensor containing $n$ affine connections.
        
        By interchanging $\mu$ and $\nu$ in Eq.\eqref{eqn4}, it is easy to see that the Riemannian connection is symmetric with respect to the two lower indices\footnote{Note that this is not a proof but a mere test, since interchange of $\mu$ and $\nu$ just have to show $\{^\alpha_{\mu\nu}\}$ to be symmetric because we made that requirement during its derivation.}. A connection of this kind is torsion-free, that is, the space admits zero torsion. However, the curvature tensor, which is derivative from the connection is nonzero\cite{gravitation73}.
        
        \subsection{Absolute Parallelism Space}
        Nonsymmetric connections are artefacts of the AP-space. The basic vector of this space is a tetrad; we denote it by $\apvec{i}{\chi}_\mu$. The construction of this vector is such that:
        \begin{equation}
        \begin{split}
        g_{\mu\nu}:=\apvec{i}{\chi}_\mu\apvec{i}{\chi}_\nu\label{eqn6}\qquad;\qquad
        \apvec{i}{\chi}^\nu\apvec{i}{\chi}_\mu=\delta^\nu_\mu\qquad;\qquad\apvec{i}{\chi}_\mu\apvec{j}{\chi}_\mu=\delta_{ij}
        \end{split}
        \end{equation}
        
        Where $\mu,\nu,i,j=0,1,2,3$. In general, since for all $\mu\neq i$, all components of $\apvec{i}{\chi_\mu}$ are distinct, $\apvec{i}{\chi}^\mu$ thus, has 16 degrees of freedom (DOF) whereas $g_{\mu\nu}$ has 10 because every $g_{\mu i}=g_{i\mu}$. The 10 DOF of $g_{\mu\nu}$ of the 4-dimensional Riemannian space is just sufficient for a field theory of gravity. Attempts have been made to unify gravity and electromagnetism using the AP-space which has 6 additional DOF to account for electromagnetic field (that is, 3 components for magnetic field and 3 components for electric field). The AP-space belongs to the Wanas \textit{type analysis} class $GIII$, which implies the AP-space has nonzero curvature tensor, nonzero energy-stress tensor and nonzero torsion\cite{Wanas2013}. Using Eq.\eqref{eqn6}, the so-called AP condition given at Eq.\eqref{eqn7} below, is easily derivable from Eq.\eqref{eqn1}.
        \begin{equation}
        \begin{split}
        0&=D_\nu\apvec{i}{\chi}_\mu\label{eqn7}
        \end{split}    
        \end{equation}
        Applying the usual rule for covariant differentiation on Eq.\eqref{eqn7} and making the arbitrary connection $\Gamma$ the subject, we have the following.
        \begin{equation}
        \begin{split}
        0&=\apvec{i}{\chi}_{\mu,\nu}-\apvec{i}{\chi}_\alpha\Gamma^\alpha{}_{\mu\nu}\\
        &=\apvec{i}{\chi}^\epsilon\apvec{i}{\chi}_{\mu,\nu}-\apvec{i}{\chi}^\epsilon\apvec{i}{\chi}_\alpha\Gamma^\alpha{}_{\mu\nu}\\
        &=\apvec{i}{\chi}^\epsilon\apvec{i}{\chi}_{\mu,\nu}-\delta^\epsilon_\alpha\Gamma^\alpha{}_{\mu\nu}\label{eqn8}
        \end{split}
        \end{equation}
        Clearly, the result of Eq.\eqref{eqn8} gives the following expression for the connection.
        \begin{equation}
        \begin{split}
        \Gamma^\epsilon{}_{\mu\nu}=\apvec{i}{\chi}^\epsilon\apvec{i}{\chi}_{\mu,\nu}\label{eqn9}
        \end{split}    
        \end{equation}
        The connection at Eq.\eqref{eqn9} is written in terms of the basic vectors of the AP-space, hence, it is called the (canonical) AP-connection. Due do the nonsymmetry of the canonical AP-connection, it should be noted that interchanging the indices $\mu$ and $\nu$ at Eq.\eqref{eqn9} yields an object different from the original and this object still possesses all the attributes of a connection as stated at Eq.\eqref{abs9}; the result is called a dual canonical AP-connection and it is as stated below.
        \begin{equation}
        \begin{split}
        \tilde{\Gamma}^\epsilon{}_{\mu\nu}=\Gamma^\epsilon{}_{\nu\mu}\label{eqn14}
        \end{split}
        \end{equation}    
        
        Also, since the canonical AP-connection at Eq.\eqref{eqn9} is nonsymmetric, it can be decomposed into nonvanishing symmetric and skew-symmetric components as shown.
        \begin{equation}
        \begin{split}
        \Gamma^\epsilon{}_{\mu\nu}={1\over2}(\Gamma^\epsilon{}_{\mu\nu}+\Gamma^\epsilon{}_{\mu\nu})+{1\over2}(\Gamma^\epsilon{}_{\mu\nu}-\Gamma^\epsilon{}_{\mu\nu})\label{eqn12}
        \end{split}
        \end{equation}
        The connection $\Gamma^\epsilon{}_{\mu\nu}$ decomposes into two parts\footnote{The decomposition of $\Gamma^\epsilon{}_{\mu\nu}$ into a tensor and a non-tensor components further accentuates its non-symmetrical nature.}: a tensor part, which is essentially one-half of the spacetime torsion as given at Eq.\eqref{eqn11} and a second part possessing all the attributes of a connection, and we identify it as yet another connection of the AP-space; it is called in literature, symmetric AP-connection.
        \begin{equation}
        \begin{split}
        \Gamma^\epsilon{}_{(\mu\nu)}:={1\over2}(\Gamma^\epsilon{}_{\mu\nu}+\Gamma^\epsilon{}_{\nu\mu})\label{eqn13}
        \end{split}
        \end{equation}
        
        To give coda on connection, we recognize that the Riemannian metric $g_{\mu\nu}$ is also admissible in the AP-space as Eq.\eqref{eqn6} indicates. And since the Riemannian metric is required to be unchanged with respect to parallel transfer, the same Riemannian connection of Riemannian space is also present in the AP-space.
        
        From the foregoing, it is clear that while the Riemannian space admits only the Riemannian connection, the AP-space admits up to four. In fact, addition of any (1, 2)-tensor of the AP-space to any AP-connection is itself a connection of the space\cite{Wanas2015}.

        
        It should be evident from Eq.\eqref{eqn9} and Eq.\eqref{abs9} that even though $\Gamma^\alpha{}_{\mu\nu}$ isn't a tensor, the difference between it and its dual, that is $\Gamma^\alpha{}_{\mu\nu}-\Gamma^\alpha{}_{\nu\mu}$, is a tensor called the torsion tensor, denoted by $\Lambda^\epsilon{}_{\mu\nu}$; evidently, it is antisymmetric with respect to the last pair or last two slots of indices. It is nonzero due to the nonsymmetry of the canonical AP-connection.
        \begin{equation}
        \begin{split}
        \Lambda^\epsilon{}_{\mu\nu}:=\Gamma^\alpha{}_{\mu\nu}-\Gamma^\alpha{}_{\nu\mu}\label{eqn11}
        \end{split}
        \end{equation}
        
        Furthermore, we state the following definition for the contortion of the AP-space.
        \begin{equation}
        \begin{split}
        \gamma^\alpha{}_{\mu\nu}=\apvec{i}{\chi}^\alpha\apvec{i}{\chi}_{\mu;\nu}\label{eqn15}
        \end{split}
        \end{equation}
        Should one perform the covariant differentiation indicated in Eq.\eqref{eqn15}, one obtains:
        \begin{equation}
        \begin{split}
        \gamma^\alpha{}_{\mu\nu}=\Gamma^\alpha{}_{\mu\nu}-\{^\alpha_{\mu\nu}\}\label{eqn16}
        \end{split}    
        \end{equation}
        And because the Riemannian connection is symmetric, it can be seen from Eq.\eqref{eqn16}, that the torsion can also be defined in terms of the contortion as follows.
        \begin{equation}
        \begin{split}
        \Lambda^\alpha{}_{\mu\nu}=\gamma^\alpha{}_{\mu\nu}-\gamma^\alpha{}_{\nu\mu}\label{eqn17}
        \end{split}    
        \end{equation}
        Since an alternative expression for contortion in terms of the torsion will arise in this text, we justify in what follows the relationship between them.
        \begin{equation}
        \begin{split}
        \begin{split}
        \Lambda^\alpha{}_{\mu\nu}&=\gamma^\alpha{}_{\mu\nu}-\gamma^\alpha{}_{\nu\mu}\\
        \Lambda^\mu{}_{\nu\alpha}&=\gamma^\mu{}_{\nu\alpha}-\gamma^\mu{}_{\alpha\nu}\\
        -\Lambda^\nu{}_{\alpha\mu}&=-\gamma^\nu{}_{\alpha\mu}+\gamma^\nu{}_{\mu\alpha}\label{eqn18}
        \end{split} 
        \end{split}
        \end{equation}
        While adding the three expressions at Eq.\eqref{eqn18}, we note that the contortion is antisymmetric with respect to the first pair or first two slots of indices: this means, switching the positions of of $\mu$ and $\nu$ in $\gamma^\nu{}_{\mu\alpha}$ makes it magnitude-equal but sign-opposite to $\gamma^\mu{}_{\nu\alpha}$. Thus, adding the lines at Eq.\eqref{eqn18}, and orienting the indices appropriately\footnote{By the way, rasing and lowering of indices is technically done by the metric tensor.}, we have that the contortion is related to the torsion as follows.
        \begin{equation}
        \begin{split}
        \gamma^\alpha{}_{\mu\nu}={1\over2}(\Lambda^\alpha{}_{\mu\nu}+\Lambda_{\mu\nu}{}^\alpha-\Lambda_\nu{}^\alpha{}_\mu)\label{eqn19}
        \end{split}
        \end{equation}
        Owing to the antisymmetric nature of the torsion tensor, we note the following results.
        \begin{equation}
        \begin{split}
        &\Lambda^\alpha{}_{\mu\nu}=-\Lambda^\alpha{}_{\nu\mu};\\
        &\forall\enspace \mu=\nu,\enspace \Lambda^\alpha{}_{\mu\mu}=-\Lambda^\alpha{}_{\mu\mu}\implies\Lambda^\alpha{}_{\mu\mu}=0;
        \end{split}
        \end{equation}
        However, in general:
        \begin{equation}
        \begin{split}
        &0\neq\Lambda^\alpha{}_{\mu\alpha}=C_\mu
        \end{split}
        \end{equation}
        Similarly:
        \begin{equation}
        \begin{split}
        \gamma^\alpha{}_{\alpha\nu}&={1\over2}(\Lambda^\alpha{}_{\alpha\nu}+\Lambda_{\alpha\nu}{}^\alpha-\Lambda_\nu{}^\alpha{}_\alpha)\\
        &={1\over2}(\Lambda^\alpha{}_{\alpha\nu}-\Lambda_{\alpha}{}^\alpha{}_\nu-0)\\&={1\over2}(\Lambda^\alpha{}_{\alpha\nu}-\Lambda^\alpha{}_{\alpha\nu}-0)\\&=0\\
        \end{split}
        \end{equation}
        However, in general:
        \begin{equation}
        \begin{split}
        \gamma^\alpha{}_{\mu\alpha}&={1\over2}(\Lambda^\alpha{}_{\mu\alpha}+\Lambda_{\mu\alpha}{}^\alpha-\Lambda_\alpha{}^\alpha{}_\mu)\\
        &={1\over2}(\Lambda^\alpha{}_{\mu\alpha}+0+\Lambda^\alpha{}_{\mu\alpha})\\
        &=\Lambda^\alpha{}_{\mu\alpha}=C_\mu    
        \end{split}
        \end{equation}
        
        In the present text, the vector $C_\mu=\Lambda^\alpha{}_{\mu\alpha}=\gamma^\alpha{}_{\mu\alpha}$ will be employed as generalized electromagnetic potential.

        To recapitulate, we give the following tabular summary of few  points highlighted so far.
        \begin{table}[H]
            \centering
            \caption{Overview of Riemannian space (RS) and AP-space}
            \begin{tabularx}{\textwidth}{|X|p{0.25\textwidth}|p{0.25\textwidth}|p{0.25\textwidth}|X|}
                \hline
                &Connections&Torsion/contortion&Basic tensor&DOF\\
                \hline
                RS&$\{^\alpha_{\mu\nu}\}$&$\Lambda^\alpha{}_{\mu\nu}=0; \gamma^\alpha{}_{\mu\nu}=0$&$g_{\mu\nu}$&$10$\\
                \hline
                &$\{^\alpha_{\mu\nu}\}$&&&\\
                AP&$\Gamma^\alpha{}_{\mu\nu}$&$\Lambda^\alpha{}_{\mu\nu}\neq0$&$\apvec{i}{\chi}^\alpha$&$16$\\
                &$\tilde{\Gamma}^\alpha{}_{\mu\nu}$&$\gamma^\alpha{}_{\mu\nu}\neq0$&&\\
                &$\Gamma^\alpha{}_{(\mu\nu)}$&&&\\
                \hline
            \end{tabularx}
        \end{table}

        In the following section, we apply a couple of the connections presented in this text in deriving equations describing the world lines of particles.

        This article is set in the following manner: the present section discusses the mathematical preliminaries of the Riemannian and the AP spaces. In \secref{two}, we present a treatment of the classical approach to path equations, precisely the Einstein-Maxwell path equations. In a subsection of \secref{three} is the Bazanski approach to path equations where the Riemannian connection is employed; in another subsection, we give an expression that is absolutely invariant along a curve in the AP-space and use it to obtain path equations involving the canonical AP-connection. Then in \secref{four}, we make some remarks and give conclusion.
        

        
        
        \section{Equations of Motions\label{two}}
        In writing the equations of motion for particles in gravitational field, we shall make use of the Euler-Lagrange equations stated below.
        \begin{equation}
        {d\over ds}{\partial L\over\partial\dot{x}^\eta}-{\partial L\over\partial x^\eta}=0\label{euler}
        \end{equation}
        As we shall see, applying the nonsymmetric connection will extend the applicability of the equations of motion and thus accounting for other features other than mere following of a geodesic by a massive particle. This section reviews the Einstein-Maxwell path equations by the usual Lagrangian method.

        \subsection{Einstein-Maxwell path Equations\label{twob}}
        If a source of gravity usually denoted as $T_{\mu\nu}$, is electromagnetic in nature, a massive particle bearing charge in the corresponding field of gravity will be described by the Einstein-Maxwell path equations. This is simply the geodesic with a forcing term giving rise to a set of inhomogeneous differential equations. The Lagrangian can be defined as follows:
        \begin{equation}
        \begin{split}
        L:=g_{\mu\nu}(\dot{x}^\mu+kA^\mu)\dot{x}^\nu\label{emq1}
        \end{split}
        \end{equation}
        Where $k$ guarantees dimensional consistency. Next, we differentiate the Lagrangian according to the required terms in the Euler-Lagrange equations.
        
        \begin{equation}
        \begin{split}
        \frac{\partial L}{\partial x^\eta}&=g_{\mu\nu,\eta}(\dot{x}^\mu+kA^\mu)\dot{x}^\nu+g_{\mu\nu}(kA^\mu_{,\eta})\dot{x}^\nu\label{emq2a}\\
        \end{split}
        \end{equation}
        \begin{equation}
        \begin{split}
        {\partial L\over\partial\dot{x}^\eta}&=g_{\mu\nu}(\dot{x}^\mu+kA^\mu)\delta^\nu_\eta+g_{\mu\nu}(\delta^\mu_\eta)\dot{x}^\nu\\
        &=g_{\mu\eta}(\dot{x}^\mu+kA^\mu)+g_{\nu\eta}\dot{x}^\nu\\
        &=g_{\mu\eta}(2\dot{x}^\mu+kA^\mu)\label{emq2b}        
        \end{split}
        \end{equation}    
        \begin{equation}
        \begin{split}
        {d\over ds}{\partial L\over\partial\dot{x}^\eta}&={d\over ds}(g_{\mu\eta})(2\dot{x}^\mu+kA^\mu)+g_{\mu\eta}{d\over ds}(2\dot{x}^\mu+kA^\mu)\\
        &={dg_{\mu\eta}\over dx^\nu}{dx^\nu\over ds}(2\dot{x}^\mu+kA^\mu)+g_{\mu\eta}(2{d\dot{x}^\mu\over ds}+k{dA^\mu\over dx^\nu}{dx^\nu\over ds})\\
        &=g_{\mu\eta,\nu}\dot{x}^\nu(2\dot{x}^\mu+kA^\mu)+g_{\mu\eta}(2\ddot{x}^\mu+kA^\mu{}_{,\nu}\dot{x}^\nu)\label{emq2c}
        \end{split}
        \end{equation}
        
        Collecting \eqref{emq2a} and \eqref{emq2c} and plugging back into the Euler-Lagrange equation \eqref{euler}, we have:
        \begin{equation}
        \begin{split}
        &g_{\mu\eta,\nu}\dot{x}^\nu(2\dot{x}^\mu+kA^\mu)\\&+g_{\mu\eta}(2\ddot{x}^\mu+kA^\mu_{,\nu}\dot{x}^\nu)-g_{\mu\nu,\eta}(\dot{x}^\mu+kA^\mu)\dot{x}^\nu-g_{\mu\nu}(kA^\mu_{,\eta})\dot{x}^\nu=0\label{emq3}
        \end{split}
        \end{equation}
        For clarity, we collect like terms of $\dot{x}^\mu\dot{x}^\nu$ and of $k\dot{x}^\nu$ and simplify them separately as follows.
        \begin{equation}
        \begin{split}
        (2g_{\mu\eta,\nu}-g_{\mu\nu,\eta})\dot{x}^\mu\dot{x}^\nu&=( 2g_{\mu\epsilon}\{^\epsilon_{\eta\nu}\}+2g_{\epsilon\eta}\{^\epsilon_{\mu\nu}\}-g_{\epsilon\nu}\{^\epsilon_{\mu\eta}\}-g_{\mu\epsilon}\{^\epsilon_{\nu\eta}\} )\dot{x}^\mu\dot{x}^\nu\\&=2g_{\epsilon\eta}\{^\epsilon_{\mu\nu}\}\dot{x}^\mu\dot{x}^\nu\label{emq4}
        \end{split}
        \end{equation}
        \begin{equation}
        \begin{split}
        k\dot{x}^\nu(g_{\mu\eta,\nu}A^\mu+g_{\mu\eta}A^\mu_{,\nu}-g_{\mu\nu,\eta}A^\mu-g_{\mu\nu}A^\mu_{,\eta})&=k\dot{x}^\nu\big((g_{\mu\eta}A^\mu)_{,\nu}-(g_{\mu\nu}A^\mu)_{,\eta}\big)\\
        &=k\dot{x}^\nu(A_{\eta,\nu}-A_{\nu,\eta})\label{emq5}
        \end{split}    
        \end{equation}
        Putting back the results of \eqref{emq4} and \eqref{emq5} into \eqref{emq3} and noting that $g_{\mu\eta}(2\ddot{x}^\mu)$ is the only term without like-terms or not collected for simplification, we have:
        \begin{equation}
        \begin{split}
        \ddot{x}^\alpha+\{^\alpha_{\mu\nu}\}\dot{x}^\mu\dot{x}^\nu&=-{1\over2}kg^{\eta\alpha}\dot{x}^\nu f_{\eta\nu}\label{emq6}
        \end{split}
        \end{equation}
        The result \eqref{emq6} differs from the geodesic only in the non-vanishing of the tensor field $f_{\eta\nu}$, which has the structure of the electromagnetic field strength tensor. The constant $k$ has been defined as charge-mass ratio of the gravitating object\cite{Lichnerowicz1955}, that is:
        \begin{equation}
        k={e\over m}\label{emq7}
        \end{equation}
        For a neutral object or for $e=0$, Eq.\eqref{emq6} will retract to the geodesic. Thus, Eq.\eqref{emq6} is said to describe the motion of a time-like (charged) object capable of interacting with the spacetime curvature as well as with the field of energy (electromagnetic in this case), which occasions the curvature.
        
        \section{Bazanki-Type Lagrangian\label{three}}
        The geodesic and even the Einstein-Maxwell path equations can also be obtained following an approach first put forward by Bazanski\cite{Kahil_Aug_2007}. 
        
        \begin{figure}[h!]
            \centering
            \includegraphics[scale=0.6]{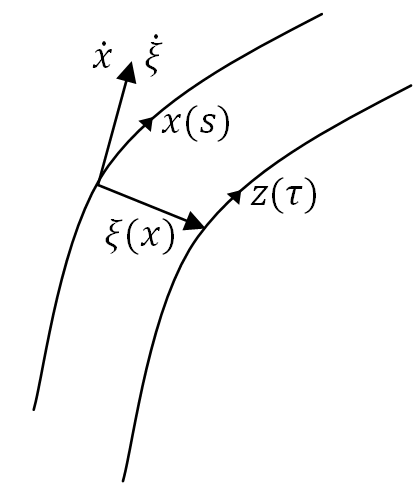}
            \caption{Deviation vector $\xi$ between parallel geodesics\label{geo_devi}}
        \end{figure}
        
        In order to obtain equations of motion by this means, the Lagrangian takes the following general form.
        \begin{equation}
        \begin{split}
        L(x,\dot{x};\xi,\dot{\xi}):=g_{\mu\nu}\dot{x}^\mu{D\xi^\nu\over ds}\label{bzk1}
        \end{split}
        \end{equation}
        
        But clearly as \figref{geo_devi} suggests, the deviation vector\footnote{Roughly speaking, the deviation vector is the vector distance between two world lines} $\xi^\nu$, has an explicit dependence on particle world line; thus, if Riemannian space is assumed, the absolute derivative of $\xi^\nu$ at Eq.\eqref{bzk1} may be expanded as:
        \begin{equation}
        \begin{split}
        L(x,\dot{x};\xi,\dot{\xi})&=g_{\mu\nu}\dot{x}^\mu\big({D\xi^\nu\over dx^\alpha}{dx^\alpha\over ds}\big)\\
        &=g_{\mu\nu}\dot{x}^\mu\big(\xi^\nu{}_{,\alpha}+\xi^\sigma\{^\nu_{\sigma\alpha}\}\big)\dot{x}^\alpha\\
        &=g_{\mu\nu}\dot{x}^\mu\big(\dot{\xi}^\nu+\xi^\sigma\{^\nu_{\sigma\alpha}\}\dot{x}^\alpha\big)
        \label{bzk2}
        \end{split}
        \end{equation}
        We recognize the velocity of the deviation vector to be a derived quantity obtained from the product $\xi^\nu{}_{,\alpha}\dot{x}^\alpha=\dot{\xi}^\nu$. Also, if an AP-space is assumed, then the covariant derivative in \eqref{bzk2} will reflect any of the AP-connections as deemed fit.
        
        We emphasize that \eqref{bzk2} is a unified Lagrangian for the geodesic as well as for the geodesic deviation; to obtain the geodesic from it, we apply variational principle with respect to the deviation vectors, and to obtain the geodesic deviation equations, we apply the variational principle with respect to the particle world line\cite{Pavsic2011}. In what follows, we derive path equations using the Bazanski-type Lagrangian with respect to  the Riemannian connections.
        
        \subsection{Equations Of Path With Riemannian Connection\label{threea}}
        Here, the vector $C^\mu$ is a geometric object as oppose to the phenomenlogical vector potential used for the Einstein-Maxwell path equations. Also, we are adopting the Bazanski-type Lagrangian. A slight variation will be observed in the resulting path equations compared to the ones in \secref{twob}. This difference, isn't from the geometry but from the methodology.       
        \begin{equation}
        L(x,\dot{x};\xi,\dot{\xi})=g_{\mu\nu}(\dot{x}^\mu+kC^\mu)(\dot{\xi}^\nu+\xi^\sigma\{^\nu_{\sigma\alpha}\}\dot{x}^\alpha)\label{rem1}
        \end{equation}
        Since we have adopted this kind of Lagrangian, to obtain equations of path, we differentiate Eq.\eqref{rem1} with respect to the deviation vector and with respect to the velocity of the deviation vector.
        \begin{equation}
        \begin{split}
        {\partial L\over\partial\xi^\epsilon}&=g_{\mu\nu}(\dot{x}^\mu+kC^\mu)\delta^\sigma_\epsilon\{^\nu_{\sigma\alpha}\}\dot{x}^\alpha\\
        &=g_{\mu\nu}(\dot{x}^\mu+kC^\mu)\{^\nu_{\epsilon\alpha}\}\dot{x}^\alpha\label{rem2}
        \end{split}
        \end{equation}
        \begin{equation}
        \begin{split}
        {\partial L\over \partial\dot{\xi}^\epsilon}&=g_{\mu\nu}(\dot{x}^\mu+kC^\mu)\delta^\nu_\epsilon\\
        &=g_{\mu\epsilon}(\dot{x}^\mu+kC^\mu)\label{rem3}
        \end{split}
        \end{equation}
        In accordance with the Euler-Lagrange equations at Eq.\eqref{euler}, next is the absolute derivative of Eq.\eqref{rem3} with respect to the evolution parameter, $s$. 
        \begin{equation}
        \begin{split}
        {d\over ds}{\partial L\over\partial\dot{\xi}^\epsilon}=g_{\mu\epsilon}(\ddot{x}^\mu+kC^\mu{}_{,\nu}\dot{x}^\nu)+g_{\mu\epsilon,\nu}\dot{x}^\nu(\dot{x}^\mu+kC^\mu)\label{rem4}
        \end{split}
        \end{equation}
        Now, we collect Eq.\eqref{rem2} and Eq.\eqref{rem4} into Eq.\eqref{euler} to write the equations of path as follows:        
        \begin{equation}
        \begin{split}
        \ddot{x}^\alpha+\{^\alpha_{\mu\nu}\}\dot{x}^\mu\dot{x}^\nu=-kC^\alpha{}_{,\nu}\dot{x}^\nu-k\{^\alpha_{\mu\nu}\}C^\mu\dot{x}^\nu\label{rem6}
        \end{split}
        \end{equation}
        To be confident that the equations are invariant with respect to parallel transportation, we will re-express the term $-kC^\alpha{}_{,\nu}\dot{x}^\nu$, in terms of covariant derivative, while noting that the metric is a constant with respect to such as a derivative. It can be shown that:        
        \begin{equation}
        \begin{split}
        -kC^\alpha{}_{,\nu}&=-kf^\alpha{}_\nu-kg^{\alpha\sigma}C_{\nu;\sigma}+kC^\epsilon\{^\alpha_{\epsilon\nu}\}\\
                          &=-kC^\alpha{}_{;\nu}+kC^\sigma\{^\alpha_{\sigma\nu}\}\label{rem7}
        \end{split}
        \end{equation}
        Hence, Eq.\eqref{rem6} may be written as given in Eqs.\eqref{rem8}\eqref{rem11}:
        \begin{equation}
        \begin{split}
        \ddot{x}^\alpha+\{^\alpha_{\mu\nu}\}\dot{x}^\mu\dot{x}^\nu=-kf^\alpha{}_\nu\dot{x}^\nu-kg^{\alpha\sigma}C_{\nu;\sigma}\dot{x}^\nu\label{rem8}
        \end{split}
        \end{equation}        
        \begin{equation}
        \begin{split}
        \ddot{x}^\alpha+\{^\alpha_{\mu\nu}\}\dot{x}^\mu\dot{x}^\nu&=-kC^\alpha{}_{;\nu}\dot{x}^\nu\label{rem11}
        \end{split}
        \end{equation}
        
         \subsection{Equations Of Path With Canonical AP-Connection\label{threeb}}
         Before we proceed, we make the assertion that along a curve in the AP-space, Eq.\eqref{nx}, is absolutely conserved with respect to any on the AP-connections.
         \begin{equation}
         L_s = \dot{x}^\alpha+kC^\alpha\label{nx}
         \end{equation}
         Instead of using a similar and lengthy approach of the previous section, we shall employ the above assertion. We shall make use of the canonical AP-connection; from Eq.\eqref{eqn16}, this connection is given by:
        \begin{equation}        
        \Gamma^\nu{}_{\sigma\alpha}=\{^\nu_{\sigma\alpha}\}+\gamma^\nu_{\sigma\alpha}\label{pcc2}        
        \end{equation}
        
        The absolute conservation of Eq.\eqref{nx} is written mathematically as follows.
        \begin{equation}
        \begin{split}
        {D\over ds}(\dot{x}^\mu+kC^\alpha)=0\label{pcc3a}
        \end{split}
        \end{equation}
        
       While noting that the tangent vector $\dot{x}^\alpha$ is dependent on the evolution parameter $s$ and the generalized electromagnetic potential, $C^\alpha$ is directly dependent on spacetime coordinates, we perform the absolute derivative of Eq.\eqref{pcc3a} as follows.


            \begin{equation}
            \begin{split}
            0=&{D\over dx^\nu}(\dot{x}^\alpha+kC^\alpha){dx^\nu\over ds}\\
            &\big[(\dot{x}^\alpha+kC^\alpha)_{,\nu}+(\dot{x}^\mu+kC^\mu)\Gamma^\alpha{}_{\mu\nu}\big]\dot{x}^\nu\\            
            &\ddot{x}^\alpha+\{^\alpha_{\mu\nu}\}\dot{x}^\mu\dot{x}^\nu+kC^\alpha{}_{,\nu}\dot{x}^\nu+kC^\mu\dot{x}^\nu(\{^\alpha_{\mu\nu}\}+\gamma^\alpha{}_{\mu\nu})+\gamma^\alpha{}_{\mu\nu}\dot{x}^\mu\dot{x}^\nu\label{pcc11proof}
            \end{split}
            \end{equation}
       
        Where in order to write the result of Eq.\eqref{pcc11proof} in a more physically useful form, we note the following.
        \begin{equation}
        \begin{split}
        C^\alpha{}_{,\nu}&=f^\alpha{}_\nu+g^{\alpha\sigma}C_{\nu\vert\sigma}-C^\mu\Lambda_{\mu\nu}{}^\alpha-C^\epsilon\Gamma^\alpha{}_{\epsilon\nu}\label{pcc12}
        \end{split}
        \end{equation}        
        Also, while noting that the torsion $\Lambda^\alpha{}_{\mu\nu}$ is antisymmetric with respect to the last pair of indices and the contortion $\gamma^\alpha{}_{\mu\nu}$ is antisymmetric with respect to the first pair of indices, we write the last term of Eq.\eqref{pcc11proof} as follows.
        \begin{equation}
        \begin{split}
        \gamma^\alpha{}_{\mu\nu}\dot{x}^\mu\dot{x}^\nu&={1\over2}(\Lambda^\alpha{}_{\mu\nu}+\Lambda_{\mu\nu}{}^\alpha-\Lambda_\nu{}^\alpha{}_\mu)\dot{x}^\mu\dot{x}^\nu\\
        &={1\over2}(\Lambda_{\mu\nu}{}^\alpha+\Lambda_{\nu\mu}{}^\alpha)\dot{x}^\mu\dot{x}^\nu\\
        &=\Lambda_{(\mu\nu)}{}^\alpha\dot{x}^\mu\dot{x}^\nu\label{pcc14}
        \end{split}
        \end{equation}
        Substituting Eq.\eqref{pcc12} and the result of Eq.\eqref{pcc14} into Eq.\eqref{pcc11proof}, we can write it in the following more physically useful form.
        \begin{equation}
        \begin{split}
        \ddot{x}^\alpha+\{^\alpha_{\mu\nu}\}\dot{x}^\mu\dot{x}^\nu=-kf^\alpha{}_\nu\dot{x}^\nu-kg^{\alpha\sigma}C_{\nu\vert\sigma}\dot{x}^\nu+kC^\mu\Lambda_{\mu\nu}{}^\alpha\dot{x}^\nu-\Lambda_{(\mu\nu)}{}^\alpha\dot{x}^\mu\dot{x}^\nu\label{pcc16}
        \end{split}
        \end{equation}
        
        \section{Remarks and Conclusion\label{four}}
       
        Eq.\eqref{pcc16} compared with the results at \secref{threea} and \secref{twob}, clearly contains many pieces of vital information. The term $-kf^\alpha{}_\nu\dot{x}^\nu-kg^{\alpha\sigma}C_{\nu\vert\sigma}\dot{x}^\nu$, represents the interaction of the generalized electrogmatic field tensor with spacetime tangent vector. Path equations of \secref{threea} involving the Riemannian connection, also contain these terms, but not the terms will shall discuss next. The term $kC^\mu\Lambda_{\mu\nu}{}^\alpha\dot{x}^\nu$, represents the interaction of tangent vector and the geometric potential with spacetime torsion. As stated in \secref{twob}, $k=e/m$, where $e$ is the charge of a particle of mass $m$. It is clear that should $e=0$, that is, for a neutral particle in AP-space, Eq.\eqref{pcc16}, reduces to\cite{Wanas_2005}:        
        \begin{equation}
        \begin{split}
        \ddot{x}^\alpha+\{^\alpha_{\mu\nu}\}\dot{x}^\mu\dot{x}^\nu=-\Lambda_{(\mu\nu)}{}^\alpha\dot{x}^\mu\dot{x}^\nu\label{sum}
        \end{split}
        \end{equation}

        Thus, every term with a factor $k$ has been associated with the electromagnetic field while the term on the RHS of Eq.\eqref{sum} can be said to represent spin-torsion interaction.
        
        In this article, we propose an expression that is absolutely conserved with respect to any of the AP-connections and we use it to obtain path equations with respect to the canonical AP-connection and we suggested physical interpretation to a term in the equations.
        
        \section*{Acknowledgement}
       We are thankful to the following: the National Mathematical Center (NMC), Abuja, the COMSATS Institute of Information Technology (CIIT), Islamabad and the Higher Education Commission (HEC) of Pakistan for jointly funding our research.
    \printbibliography
\end{document}